\documentclass[preprint,preprintnumbers,amsmath,amssymb]{revtex4}

\usepackage{graphicx}
\usepackage{dcolumn}
\usepackage{bm}

\begin{document}

\preprint{preprint}

\title{How a single stretched polymer responds coherently to a minute oscillation in fluctuating environments: An entropic stochastic resonance}

\author{Won Kyu Kim}
\author{Wokyung Sung}%
 \thanks{Corresponding author} 
  \email{wsung@postech.ac.kr}
\affiliation{%
Department of Physics and POSTECH Center for Theoretical Physics, Pohang University of Science and Technology (POSTECH), Pohang, 790-784, Republic of Korea\\
}

\begin{abstract}
Within the cell, biopolymers are often situated in constrained, fluid environments, e.g., cytoskeletal networks, stretched DNAs in chromatin. It is of paramount importance to understand quantitatively how they, utilizing their flexibility, optimally respond to a minute signal, which is, in general, temporally fluctuating far away from equilibrium. To this end, we analytically study viscoelastic response and associated stochastic resonance (SR) in a stretched single semi-flexible chain to an oscillatory force or electric field. Including hydrodynamic interactions between chain segments, we evaluate dynamics of the polymer extension in coherent response to the force or field. We find power amplification factor of the response at a noise-strength (temperature) can attain the maximum that grows as the chain length increases, indicative of an entropic stochastic resonance (ESR). In particular for a charged chain under an electric field, we find that the maximum also occurs at an optimal chain length, a new feature of ESR. The hydrodynamic interaction is found to enhance the power amplification, representing unique polymer cooperativity which the fluid background imparts despite its overdamping nature. For the slow oscillatory force, the resonance behavior is explained by the chain undulation of the longest wavelength. This novel ESR phenomenon suggests how a biopolymer self-organizes in an overdamping environment, utilizing its flexibility and thermal fluctuations.
\end{abstract}

\maketitle

\section{\label{sec:level1}Introduction}

Within the cell, the biopolymers such as DNA, proteins, and actin filaments are confined subject to various kinds of forces or fields.
In chromosome, for example, double stranded (ds) DNA is condensed in a fascinating hierarchy of varying structures, interwound via histone proteins subject to tension in chromatin level \cite{luger}.
Associated with a cellular membrane, cytoskeletal network consists of actin filaments, upon which myosins bind to exert tension.
An universal feature that underlies in these systems over various levels is the polymer flexibility that can adapt to ambient thermal noises as well as the external forces.
Another one is the aqueous background which transmits hydrodynamic interactions (HI).
In this vein, we pose a fundamental question : How the nature allows the confined (stretched) biopolymer systems to utilize the ambient thermal fluctuations and the system flexibility for maximum cooperativity to an external signal in such an overdamping media?

The stochastic resonance (SR) is a noise-induced phenomenon, where the background noise can enhance coherence and resonance of a non-linear system to a small additive periodic signal \cite{sr1,sr2}. 
The SR is a measure of coherence, synchrony, and sensitivity of the system to external influences.
The exemplar model is a bistable system of a single particle that undergoes interwell hopping transition coherent to the signal at an optimal noise-strength \cite{sr1,sr2}.
Also, the SR is recently found to occur even in a non-linear monostable system of single particle or single degree of freedom \cite{0295-5075-65-1-007,0253-6102-51-2-19}.

As linearly interconnected systems they are, polymers manifest unusual cooperativity and collective dynamics in response to those external stimuli.
It has recently been shown that the chain interconnectivity amplifies the thermally-induced hopping and the associated SR of a flexible polymer moving under a bistable potential \cite{asfaw,PhysRevE.63.021115}. The SR for such polymeric systems is of entropic nature, which shows the maximal coherence to an oscillatory force not only at optimal noise-strengths, but also, remarkably, at the optimal chain lengths that depend on chain flexibility and conformational changes. This entropic SR suggests novel possibilities for manipulating the dynamics of single biopolymer 
transitions utilizing their flexibility.

To address the SR of biopolymers in constrained situations mentioned earlier, in this work we study the dynamics of an end-tethered worm-like chain (WLC) subject to a constant tension and an additional small oscillating force.
The end-to-end distance (extension) of the chain is indeed found to respond most coherently at an optimal noise-strength, and, at optimal length for the case of the polyelectrolyte under an electric field.
It turns out that the most pronounced SR is due to the longest-wave length lateral undulation of the chain in response to the oscillation.
We show and discuss how these system parameters and HI in water with temperature-dependent viscosity affect the SR behavior.
Eventually, the effective dynamics for the chain extension is governed by the Langevin equation subject to a non-linear monostable free energy due to the constant tension force and entropic restoring force, as well as a non-constant drag coefficient that incorporates HI and chain inextensibility.

Sec.II is a brief introduction of the dynamics of a stretched semi-flexible chain in linear response to a time dependent force, including HI.
In Sec.III, we specialize in the response of the chain extension, the SR in particular, to a small oscillating force applied to a chain end.
SR is evaluated as a function of the noise-strength as well as chain parameters and HI.
In Sec.IV, we extend the analytical results of Sec.III to the case of a charged polymer under an oscillating electric field.
In Sec.V, to analyze the nature of entropic SR and effect of HI, the chain dynamics is projected into one-dimensional dynamics of the extension under a free energy and an effective drag.

\section{Dynamics of semi-flexible chain under a time-dependent tension}

\begin{figure}
\includegraphics[width=8.6cm]{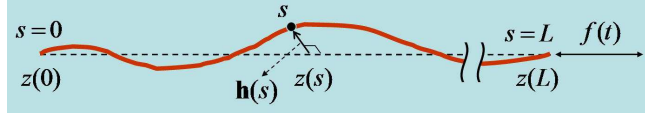}
\caption{\label{fig:system} The coordinates of the semi-flexible chain in a fluid.
}
\end{figure}
Consider a semi-flexible chain whose one end is pulled by a time-dependent force. Provided that the force is strong enough to stretch the chain nearly to the full contour length $L$, we represent the position vector of a segment at an arc length $s$ in a cylindrical coordinates ${\bf r}(s,t)=({\bf h}(s,t),z(s,t))$ where ${\bf h}(s,t)$ is radial undulation vector and $z(s,t)$ is arc length projected along the force (Fig.~\ref{fig:system}).

Due to segment inextensibility, $|\frac{\partial {\bf r}(s)}{\partial s}|=1$, and small undulation, $\left(\frac{\partial {\bf h}(s)}{\partial s}\right)^{2}\ll1$, we consider here, the chain extension along the force is
\begin{equation}
Z(t)=\int_{0}^{L}~ds \left(\frac{\partial z}{\partial s}\right)\approx L-\frac{1}{2}\int_{0}^{L}~ds \left(\frac{\partial {\bf h}(s)}{\partial s}\right)^{2}.
\label{eq:1}
\end{equation}
Of our major interest is how the average end-to-end distance, $\langle Z(t)\rangle$, responds to a time-dependent force denoted by $f(t)$, depending on the chain parameters (such as stiffness and length) and on the fluid environment characteristics (such as the temperature and HI).
In this study we confine ourselves to the force $f(t)={f}_{0}+\delta{f}(t)$ where $|\delta{f}(t)|\ll{f}_{0}$. Initially(at $t=0$), the chain is brought to a thermal equilibrium under a constant and uniform force ${f}_{0}$.
The dynamic response of the mean chain extension to a time-dependent force in a viscous liquid was addressed in references \cite{granekprl,ohta}.
For logical flow we recapitulate here the basic equations and obvious expressions that were already derived or known.
However, a modification is made in the chain mobility by incorporating HI as well as self-frictions.
This is crucial for assessing the effect of HI on SR in our study.

To study the dynamics of chain extension afterwards, we consider an effective Hamiltonian for the chain pulled by the force given by
\begin{eqnarray}
\mathcal{H}&=&\frac{1}{2}\int_0^L ds \kappa\left( {\frac{{\partial^2
{\bf{h}}(s)}}{{\partial s^2 }}} \right)^2-f(t)z(L,t)\nonumber\\
&=&\frac{1}{2}\int_0^L ds\left[\kappa\left( {\frac{{\partial^2
{\bf{h}}(s)}}{{\partial s^2 }}} \right)^2+f(t)\left( {\frac{{\partial
{\bf{h}}(s)}}{{\partial s}}} \right)^2\right]\nonumber\\
&&-f(t)\left(L-z(0,t)\right),
\label{eq:2}
\end{eqnarray}
where two terms in the $[~]$ represent chain bending and stretching energy, and the last term is trivial with $z(0,t)=0$. $\kappa$ is the bending rigidity of the chain and is related to the chain persistence length $L_{p}$ by $\kappa=k_{B}T L_{p}$.
The dynamics of the ${\bf{h}}(s,t)$ is given by the Langevin equation
\begin{eqnarray}
\frac{\partial {\bf h}(s,t)}{\partial t}=\int_{0}^{L}ds'{\bf\Lambda}(s,s')\cdot{\bf F}(s',t)+{\bf{R}}(s,t).
\label{eq:lang}
\end{eqnarray}
Here, ${\bf\Lambda}(s,s')=\frac{\bf 1}{\zeta}\delta(s-s')+\frac{1}{8 \pi \eta |{\bf r}_{ss'}|}\left[{\bf 1}+\frac{{\bf r}_{ss'}{\bf r}_{ss'}}{|{\bf r}_{ss'}|^2}\right]$ is the mobility tensor consisting of self friction (with the friction coefficient $\zeta$ per unit length of the chain) and hydrodynamic interaction between two different segments $s$ and $s'$, at a distance ${\bf r}_{ss'}\equiv{\bf r}(s)-{\bf r}(s')$, where $\bf{1}$ is unit tensor, $\eta$ is the fluid viscosity.
The ${\bf{R}}(s,t)$ is the Gaussian and white noise. ${\bf F}(s',t)$ is the density of the radial force acting on segment $s'$,
\begin{eqnarray}
{\bf{F}}(s',t)&=&-\delta \mathcal{H}/\delta {\bf{h}}(s',t)\nonumber\\
&=&-\kappa\left( {\frac{{\partial^4
{\bf{h}}}}{{\partial s'^4 }}} \right)+f(t)\left(\frac{{\partial^2 {\bf{h}}}}{{\partial s'^2}}\right).
\end{eqnarray}

Within the small undulation approximation we consider, the ${\bf\Lambda}(s,s')={\bf 1}\left[\delta(s-s')/{\zeta}+{1}/{8 \pi \eta |s-s'|}\right]$ to reduce the Eq.~(\ref{eq:lang}) as
\begin{eqnarray}
\frac{\partial {\bf h}(s,t)}{\partial t}=\zeta^{-1}{\bf F}(s,t)+\int_{0}^{L}ds'\frac{{\bf F}(s',t)}{8 \pi \eta |s-s'|}+{\bf{R}}(s,t).
\label{eq:lang2}
\end{eqnarray}
By the Fourier transform, e.g., ${\bf h}(q,t)=\int_{-\infty}^{\infty}ds~e^{i q s}~{\bf h}(s,t)$, the Eq.~(\ref{eq:lang2}) is rewritten \cite{granekprl} as
\begin{equation}
\frac{\partial {\bf h}(q,t)}{\partial t}=-\omega(q,t){\bf h}(q,t)+{\bf{R}}(q,t),
\label{eq:eom3}
\end{equation}
where $\omega(q,t)=\Lambda(q)\left[\kappa q^4+f(t) q^2\right]$, $\Lambda(q)=1/{\zeta}+K_{0}(qa)/4\pi\eta$, and $K_{0}(qa)$ is the modified Bessel function of the second kind with a cutoff $a$ introduced as the chain diameter \cite{PhysRevLett.71.1864}.
The average chain extension is given by
\begin{equation}
\frac{\langle Z(t)\rangle}{L}=1-\frac{1}{2 L^2}\sum_{q=\pi/L}^{q=\pi/a}q^{2}\langle{\bf h}(q,t)\cdot{\bf h}(-q,t)\rangle.
\label{eq:ete5}
\end{equation}
From Eq.~(\ref{eq:eom3}), we obtain to the order linear in $\delta f$,
\begin{eqnarray}
\langle{\bf h}(q,t)\cdot{\bf h}(-q,t)\rangle=\langle{\bf h}(q,0)\cdot{\bf h}(-q,0)\rangle-\int\limits_0^t {dt'}m(q,t-t')\delta f(t'),\nonumber
\end{eqnarray}
where $\langle{\bf h}(q,0)\cdot{\bf h}(-q,0)\rangle={{2k_B TL}}/{{\left( {\kappa q^4  + f_0 q^2 } \right)}}$ and
\begin{equation}
m(q,t-t')=\frac{{4 k_B TL\Lambda(q)}}
{{\left( {\kappa q^2  + f_0 } \right)}}e^{-\frac{t-t'}{\tau_{q}}}.\nonumber
\end{equation}
The
\begin{equation}
\tau_{q}=\left[2\omega(q,0)\right]^{-1}=1/\left[2\Lambda(q)\left( {\kappa q^4  + f_0 q^2 } \right)\right]
\label{eq:mm}
\end{equation}
is the relaxation time of undulation mode $q$ stretched by the force $f_{0}$. At $t=0$, Eq.~(\ref{eq:ete5}) yields the well-known relation between equilibrium stretching force and extension $Z_{0}/L\equiv\langle Z(0)\rangle/L=1-k_{B}T/\sqrt{4\kappa f_{0}}$ \cite{marko} or
\begin{equation}
f_{0}=\frac{k_{B}T}{4 L_{p}}\left(1-\frac{Z_{0}}{L}\right)^{-2}.
\label{eq:10}
\end{equation}
This is due to entropy change associated with the extension. The Eq.~(\ref{eq:ete5}) is rewritten as
\begin{eqnarray}
\langle \delta Z(t)\rangle&\equiv&\langle Z(t)\rangle-Z_{0}\nonumber\\
&=&\int\limits_0^t {dt'}M(t-t')\delta f(t'),
\label{eq:ete}
\end{eqnarray}
in terms of a dynamic response function,
\begin{eqnarray}
M(t-t')&=&{\sum_{q~}^{~}} q^{2} m(q,t-t')\nonumber\\
&=&\frac{1}{2 \pi}\int_{\pi/L}^{\pi/a} dq~q^{2}\frac{{4 k_B TL\Lambda(q)}}
{{\left( {\kappa q^2  + f_0 } \right)}}e^{-\frac{t-t'}{\tau_{q}}}.
\nonumber\\
\label{eq:dynres}
\end{eqnarray}

The Eq.~(\ref{eq:ete}) and Eq.~(\ref{eq:dynres}) represent the chain extension in linear response to time-dependent tension of general form in a viscous fluid under HI. We mention here that these equations were used in relaxation dynamics of the chain extension under time-dependent drag exerted by an end-anchored large bead \cite{granekprl} and also in viscoelastic dynamics under oscillatory force without HI \cite{ohta}, respectively.
Our study here differs from theirs in the explicit expressions for the mobility $\Lambda(q)=1/\zeta+K_{0}(qa)/4\pi\eta$, and also in the approximation via a linear response theory described next.

Although we will analyze Eq.~(\ref{eq:dynres}) in detail for SR, we need to obtain its approximation that allows us to gain much simpler analytical understanding.
We first note that the Eq.~(\ref{eq:ete}) is the consequence of a dynamic linear response theory, according to which,
\begin{equation}
M(t)=-\frac{1}{k_{B}T}\frac{d}{dt}\langle\delta Z(t) \delta Z(0)\rangle_{0},
\end{equation}
where $\langle ... \rangle_{0}$ means the average over the equilibrium ensemble in the absence of $\delta f(t)$.
\begin{figure}
\begin{center}
\includegraphics[width=8.6cm]{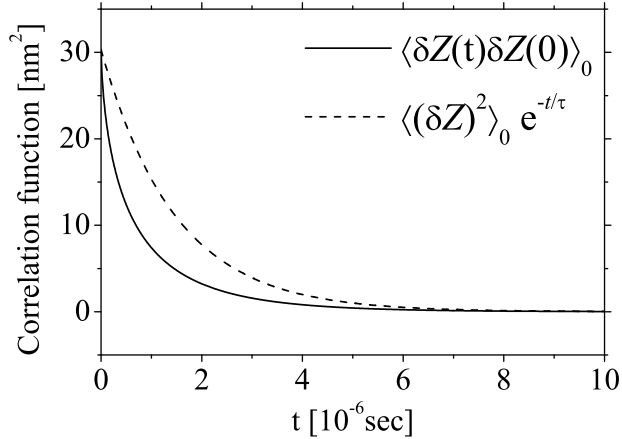}
\caption{\label{fig:corr} The correlation function of the chain extension, $\langle\delta Z(t) \delta Z(0)\rangle_{0}$ (solid curve). We use $T=37^{\circ}$C, $L=100$ nm, $L_{p}=50$ nm, $a=2$ nm and $f_{0}=0.5$ pN. The single relaxation time approximation $\langle\delta Z(t) \delta Z(0)\rangle_{0}\simeq\langle (\delta Z)^{2}\rangle_{0}e^{-t/\tau}$ is depicted by a dashed curve.}
\end{center}
\end{figure}
Integrating the above relation, we obtain the time correlation function
\begin{eqnarray}
\langle\delta Z(t) \delta Z(0)\rangle_{0}&=&-k_{B}T\int_{0}^{t}dt'M(q,t-t')+\langle (\delta Z)^{2}\rangle_{0}\nonumber\\
&=&\frac{(k_{B}T)^{2}L}{\pi}\int_{\pi/L}^{\pi/a}dq\frac{1}{\left( {\kappa q^2  + f_0 } \right)^2}e^{-\frac{t}{\tau_{q}}},\nonumber\\
\label{eq:20}
\end{eqnarray}
which at $t=0$ is reduced to equilibrium fluctuation,
\begin{equation}
\langle (\delta Z)^{2}\rangle_{0}=\frac{(k_{B}T)^{2}L}{\pi}\int_{\pi/L}^{\pi/a}dq\frac{1}{\left( {\kappa q^2  + f_0 } \right)^2}.
\label{eq:21}
\end{equation}

Fig.~\ref{fig:corr} depicts the relaxation behavior of the correlation function (Eq.~(\ref{eq:20})) of a chain stretched by the force $f_{0}$ for the parameters typical in stretched DNA. The correlation decays faster in time as the stretching force or the bending rigidity increase.
In the long time regime, the correlation function is well approximated as
\begin{equation}
\langle\delta Z(t) \delta Z(0)\rangle_{0}\simeq\langle (\delta Z)^{2}\rangle_{0}e^{-t/\tau},
\label{eq:apprco}
\end{equation}
where $\tau$ is the longest relaxation time associated with the mode of a longest wavelength ($q=\pi/L$) undulation given by
\begin{equation}
\tau=\frac{1}{2\Lambda(\pi/L)\left( {(\pi/L)^4\kappa+(\pi/L)^2f_{0} } \right)}.
\label{eq:longtau}
\end{equation}
This single relaxation time approximation is valid in the long time limit as well as at $t=0$, so that it can incorporate exactly the response to the static and low-frequency forces, although it yields a rather slowly decaying correlation (Fig.~\ref{fig:corr}).
In this approximation $M(t)$ is given as 
\begin{equation}
\tilde{M}(t)=\frac{\langle(\delta Z^2)\rangle_{0}}{k_{B}T}\frac{1}{\tau} e^{-t/\tau},
\label{eq:mtau}
\end{equation}
in a form much simpler than Eq.~(\ref{eq:dynres}).

\section{The entropic stochastic resonance of chain extension under an oscillating force on an end}

Consider a time-dependent force in the form, $\delta f(t)=f_{m} \sin(\Omega t)$, to be applied  at one end of the chain while the other end is held fixed. Since we assume that the tension is uniformly acting along the chain, we should consider the cases where the tension propagation time, $\tau_f=\eta L^2 \sqrt{\kappa}/(25L_{p} f_{0}^{3/2})$ \cite{epl47} should be much shorter than the time scales of the $f(t)$ and the $\langle \delta Z(t) \rangle$.
From the Eq.~(\ref{eq:ete}), the long time asymptotic behavior of the chain extension becomes
\begin{equation}
\langle \delta Z(t)\rangle=f_{m}\left[\chi'\sin(\Omega t)-\chi''\cos(\Omega t)\right],
\label{eq:ete2}
\end{equation}
where
\begin{eqnarray}
\chi'&=&\frac{{4k_B T L}}
{\pi }\int_{\pi/L}^{\pi/a} {dq} \frac{{q^4 \Lambda ^2 (q)}}
{{\tau_{q}^{-2} + \Omega^2 }},\nonumber\\
\chi''&=&\frac{{4k_B T L}}
{\pi }\int_{\pi/L}^{\pi/a} {dq} \frac{{\Omega\tau_{q}q^4 \Lambda ^2 (q)}}
{{\tau_{q}^{-2}  +\Omega^2 }}.\nonumber\\
\label{eq:ete3}
\end{eqnarray}
The Eq.~(\ref{eq:ete2}) is rewritten as
\begin{equation}
\left\langle\delta {Z(t)} \right\rangle  = |\chi(\Omega)| f_{m} \sin \left( {\Omega t - \varphi } \right),
\label{eq:ete4}
\end{equation}
where $|\chi|=\sqrt {\chi'^2  + \chi''^2 }$ is the absolute magnitude of dynamic susceptibility $\chi(\Omega)$ and $\varphi=\tan^{-1}\left(\chi''/\chi'\right)$ is the phase delay. The power amplification factor, $P\equiv|\chi|^{2}$, is a measure of coherence or synchrony with phase delay to the oscillatory force.

Although we will study the above $P$ for SR behavior, its single relaxation time approximation $\tilde{P}$ is useful.
Noting that the dynamic susceptibility $\chi(\Omega)$ is nothing but the Fourier-Laplace transform of $M(t)$, $\tilde{\chi}(\Omega)=\int_{0}^{\infty}dt~e^{i\Omega t} \tilde{M}(t)$,
\begin{eqnarray}
\tilde{\chi}(\Omega)=\frac{\langle (\delta Z)^{2}\rangle_{0}}{k_{B}T}\frac{1}{1-i\Omega\tau},
\label{eq:chiomega}
\end{eqnarray}
which, in the limit $\Omega\rightarrow 0$, recovers the static susceptibility of the chain $\chi_{0}=\langle (\delta Z)^{2}\rangle_{0}/{k_{B}T}$ in response to the static force $f_{0}$.
We then obtain the power amplification factor in this approximation,
\begin{eqnarray}
\tilde{P}=\left(\frac{\langle (\delta Z)^{2}\rangle_{0}}{k_{B}T}\right)^2\frac{1}{1+(\tau\Omega)^2}.
\label{eq:chisq}
\end{eqnarray}

We examine characteristics of the power amplification factor $P$ for a semi-flexible chain taking the parameter values of a double stranded DNA (dsDNA) fragments, $a=2$ nm, $L_{p}=50$ nm and $20\leq L\leq125$ nm.
In addition, we incorporate temperature dependence of the water viscosity $\eta$ given by the empirical relation $\log \frac{\eta}{\eta_{0}}=\frac{{a_1}(20-T)-{a_2}(20-T)^2}{T+{a_3}}$ where $\eta_{0}=1$ cP is the viscosity at $T=20^{\circ}$C, $a_1 =1.1709$, $a_2 =0.001827$, $a_3 =89.93$ and $T$ is temperature in the Celsius scale \cite{doi:10.1021/j100721a006}.
A chain with $L=100$ nm and $L_p =50$ nm pulled by $f_{0}=0.5$ pN in water at $T=310$ K $=37^{\circ}$C is stretched by $0.86L \leq \langle Z \rangle \leq 0.96L$, according to the Eq.~(\ref{eq:10}). 
\begin{figure*}
\begin{center}
\includegraphics[width=16.5cm]{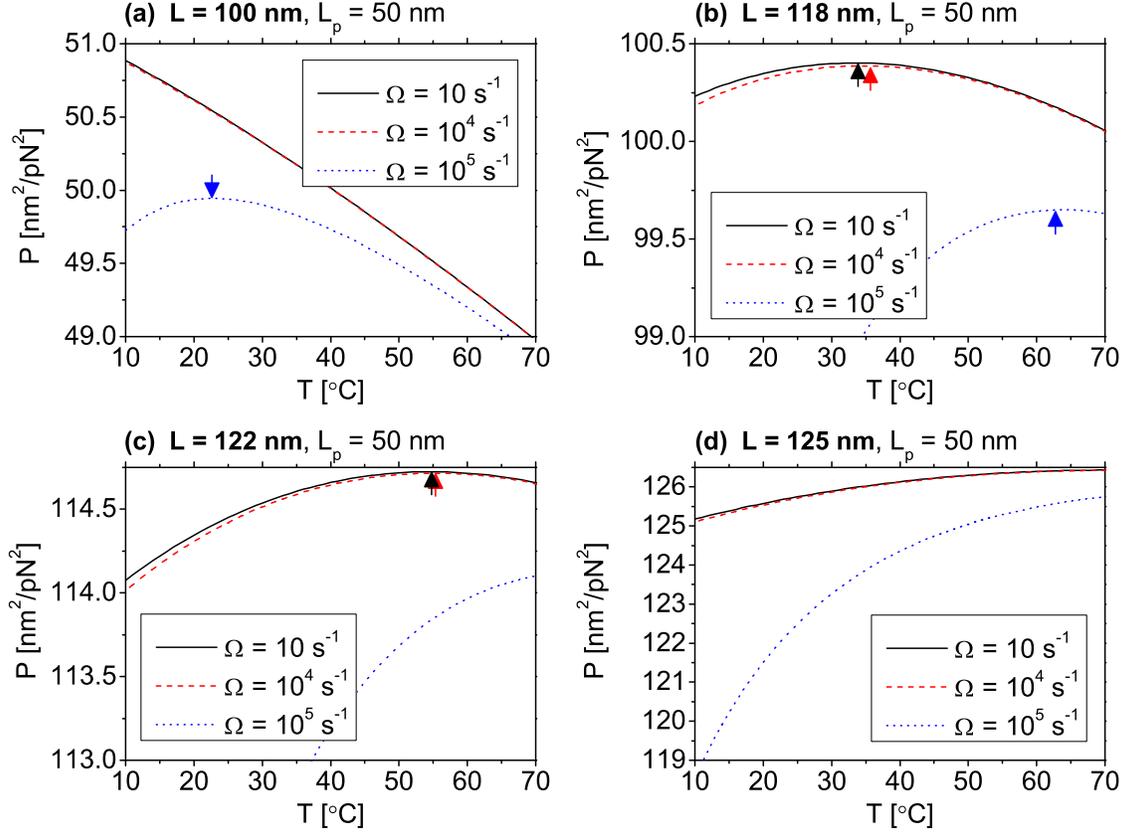}
\caption{\label{fig:2} The power amplification factor $P$ vs. the temperature in Celsius for three frequencies $\Omega=10$ s$^{-1}$ (solid), $\Omega=10^{4}$ s$^{-1}$ (dashed) and $\Omega=10^{5}$ s$^{-1}$ (dotted), with the hydrodynamic interactions (HI) included. For the semi-flexible chain with dsDNA parameters, $a=2$ nm, $L_{p}=50$ nm, and $f_{0}=0.5$ pN, the $P$ are maximized at optimal temperatures indicated by the arrows, for contour lengths (a)$L=100$ nm, (b)$L=118$ nm, (c)$L=122$ nm, and (d)$L=125$ nm.}
\end{center}
\end{figure*}
Fig.~{\ref{fig:2}} depicts the $P(T)$ with four different contour lengths $L=100$, 118, 122 and 125 nm, respectively.
The angular frequencies $\Omega$ of the driving force are chosen to be smaller than the tension propagation frequency $\Omega_{f}=2\pi/\tau_{f}$, which is about $\sim 10^{7}$ s$^{-1}$ for the parameters we consider.

As indicated by arrows in the Fig.~{\ref{fig:2}}, the $P(T)$ for a semi-flexible chain (with parameters considered) the dsDNA can have peaks with respect to the noise-strength $T$, which is the main characteristics of the SR, for the certain ranges of polymer parameter $L$ and angular frequency $\Omega$.
In one-dimensional picture of the chain dynamics which will be described in Sec.V, the chain extension is subject to an effective non-linear potential or the free energy, which depends explicitly on temperature due to the entropy of the chain.
Consequently, the SR obtained here is of the entropic nature and is dubbed as the Entropic Stochastic Resonance (ESR).

\begin{figure*}
\begin{center}
\includegraphics[width=16.5cm]{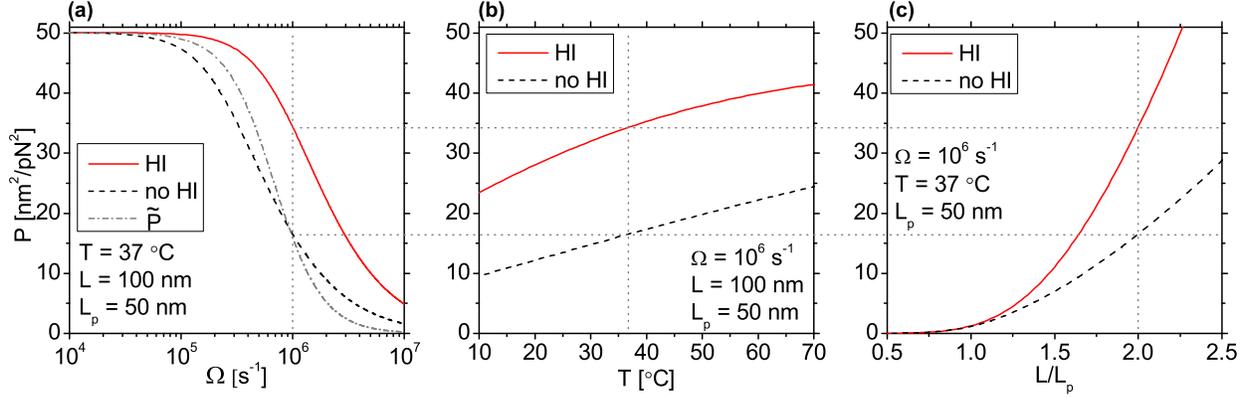}
\caption{\label{fig:4} Enhancement of the power amplification $P$ by HI. (a) $P$ vs. $\Omega$ with (solid curve) and without HI (dashed curve), at $T=37^{\circ}$C and $L=100$ nm. HI can enhance the amplification about twice at $\Omega=10^{6}$ s$^{-1}$ (see dotted vertical lines). The power amplification $\tilde{P}$ by the single relaxation time approximation is depicted by the dash-dotted curve. (b) $P$ vs. $T$ with and without HI, at $\Omega=10^{6}$ s$^{-1}$ and $L=100$ nm. (c) $P$ vs. $L/L_{p}$ with and without HI, at $T=37^{\circ}$C and $\Omega=10^{6}$ s$^{-1}$. Each cross point of the dotted lines depicts the $P$ at $\Omega=10^{6}$ s$^{-1}$, $T=37^{\circ}$C and $L=100$ nm. All the other parameters are the same as those used in the Fig.~{\ref{fig:2}}.}
\end{center}
\end{figure*}

The $P(T)$ behavior in the Fig.~{\ref{fig:2}} varies sensitively to small variations of $L$. For example, the solid curves (the $P(T)$ with the lowest angular frequency $\Omega=10 ~\text{s}^{-1}$) in the Fig.~{\ref{fig:2}}(a), (b), (c) and (d) are drastically distinct from one another only with difference of $L$ by $\sim 10$ nm; they monotonically decreasing, non-monotonically varying with a peak, and monotonically increasing, respectively, within the range of temperature shown in the Fig.~{\ref{fig:2}}.
In addition, each $P(T)$ in the Fig.~{\ref{fig:2}}(a), (b), (c) and (d) changes greatly as $\Omega$ decreases from $10^5$ s$^{-1}$ to $10^4$ s$^{-1}$, while it negligibly changes as $\Omega$ decreases below $10^4$ s$^{-1}$, converging rapidly to the $\Omega\rightarrow 0$ limit, which is explained below and seen in Fig.~\ref{fig:4}(a) later.
We find that the response of a 118 nm-long chain with dsDNA parameters pulled by 0.5 pN force and driven by a minute oscillation at $\Omega=10^{4}$ s$^{-1}$ is maximized at an optimal noise-strength which is around the physiological temperature, 37$^{\circ}$C (dashed curve in the Fig.~{\ref{fig:2}}(b)).
Of particular note is that the SR peak is quite broad (rather than feeble) within the temperature range where the water can exist in the liquid state, suggesting that a biopolymer in such condition can be widely responsive to a minute external force.

If $\Omega \ll \tau^{-1}$, the power amplification factors, coached by Eq.~(\ref{eq:chisq}), become independent of dynamics details and are largely determined by the static susceptibility $\chi_{0}=\langle (\delta Z)^{2}\rangle_{0}/k_{B}T$, from which we can evaluate the conditions for SR peaks.
Performing the integration Eq.~(\ref{eq:21}) for the cases where $f_{0}{L^2}/(k_{B} T L_{p})\gg 1$,
\begin{equation}
\chi_0 \approx \frac{L}{4 L_{p}^{2}}\left(\frac{k_{B} T L_{p}}{f_{0}}\right)^{3/2}-\left(\frac{k_{B}T}{f_{0}}\right)^2,
\label{eq:26}
\end{equation}
leading to the optimal temperature for SR,
\begin{equation}
T^{\ast}=\frac{9}{256}\frac{f_{0}L^2}{k_{B}L_{p}}.
\label{eq:27}
\end{equation}
Among others, this shows that $T^{\ast}$ increases (sharply) with $L$, in consistency with the peaks at low frequencies in Fig. {\ref{fig:2}}(b) and (c), as mentioned earlier.
It establishes an important condition for SR; the chain does not have a SR if $f_{0}=0$, i.e., unless the chain is constrained. Also, it indicates that neither an ideal chain ($L_{p}=0$) nor a rigid chain ($L_{p}=\infty$) can support the SR.

Fig.~\ref{fig:4} shows enhancement of the power amplification $P(\Omega,T,L/L_{p})$ by HI.
The Fig.~\ref{fig:4}(a) depicts $P$ vs. $\Omega$ with and without HI, at $T=37^{\circ}$C and $L=100$ nm.
With HI turned off, $\Lambda(q)\rightarrow1/\zeta$ with $\zeta$ put as ${3\pi\eta}$, the $P$ is reduced to about half that at $\Omega=10^{6}$ s$^{-1}$, meaning that HI enhances the $P$.
Note that the effect of HI is not significant for the low angular frequencies below $\tau^{-1}$, which is about $10^5$ s$^{-1}$ for the parameters considered over which the chain segmental dynamics associated with HI is rapidly relaxed out and the static response ($\Omega \rightarrow 0$) dominates.
This feature is explained by the single relaxation time approximation $\tilde{P}$, which is depicted by the dash-dotted curve in Fig~\ref{fig:4}(a).
It evidently become an excellent approximation to $P$ as $\Omega$ decreases to zero.
Since the relaxation time $\tau$ decreases with HI as shown by Eq.~(\ref{eq:longtau}), the Eq.~(\ref{eq:chisq}) indicates SR is enhanced by HI for $\Omega \gtrsim \tau^{-1}$.
Since HI makes $\tau^{-1}$ larger by the factor $\sim K_{0}(\pi a/L)\approx|\ln(\pi a/L)|$ compared with that without HI, HI effect can be greatly enhanced for longer contour length, as shown by Fig. \ref{fig:4}(c).
The $P$ at fixed $\Omega=10^{6}$ s$^{-1}$, $T=37^{\circ}$C and $L=100$ nm with and without HI are depicted as cross points of the dotted lines, in the Fig.~\ref{fig:4}.

\section{The entropic stochastic resonance of chain extension under an oscillating electric field}

In this section, we consider a polyelectrolyte chain which is uniformly charged with a linear charge density $\lambda$, subject to a time dependent electric field ${\bf E}(t)=-\left[E_{0}+\delta E(t)\right]{\bf \hat{z}}$ with $\delta E \ll E_{0}$. The effective Hamiltonian is then \cite{benetatos,maier,hori},
\begin{equation}
\mathcal{H}=\int_0^L {ds \left[\frac{\kappa}{2}\left( {\frac{{\partial^2
{\bf{h}}(s)}}{{\partial s^2 }}} \right)^2-{E}(t)\lambda Z(s,t)\right]}.
\label{eq:He0}
\end{equation}
An effect of electrostatic interactions between the segments are incorporated into a renormalization of the persistence length \cite{OSF1,OSF2} that depends upon the salinity of the background fluid. The perturbation effect of the $\delta E(t)$ on the Hamiltonian is
\begin{equation}
\mathcal{H'}=-L \lambda \delta E(t)Z_{\text{cm}}(t),
\label{eq:He2}
\end{equation}
where $Z_{\text{cm}}(t)=\int_{0}^{L} ds Z(s,t)/L$ is the chain center of mass position along the field. Hence, unlike the previous case of an end-stretching force, the $Z_{\text{cm}}(t)$ is the coordinate conjugates to the force -$L \lambda \delta E(t)$ acting on entire segments.
The mean of $Z_{\text{cm}}(t)$ in the small undulation limit is shown to be half the mean end-to-end distance.
According to phenomenological and analytical arguments, equilibrium end-to-end distance under the strong electric field is \cite{marko,maier,hori},
\begin{equation}
\frac{Z_{0}}{L}=1-\sqrt{\frac{k_{B}T}{L_{p}L |\lambda|E_{0}}},
\label{eq:eteE}
\end{equation}
which means that the uniform electric force corresponds to the end-stretching force via the relation $f_{0}= L|\lambda| E_{0}/4$. The response of the mean center of mass position to the oscillating electric field $\delta E(t)=\delta E \sin\left(\Omega t\right)$ is then obtained by replacing $f_{m}$ by $L |\lambda|\delta E/4$ and $\chi$ by $\chi/2$ in the Eq.~(\ref{eq:ete3}) and Eq.~(\ref{eq:ete4}). The power amplification factor then becomes $P_{E}=|\chi|^{2}/4$ with $f_{0}$ replaced by $L|\lambda| E_{0}/4$. Note that the electric field induces a tension which is proportional to the chain length $L$.

\begin{figure*}
\begin{center}
\includegraphics[width=16.5cm]{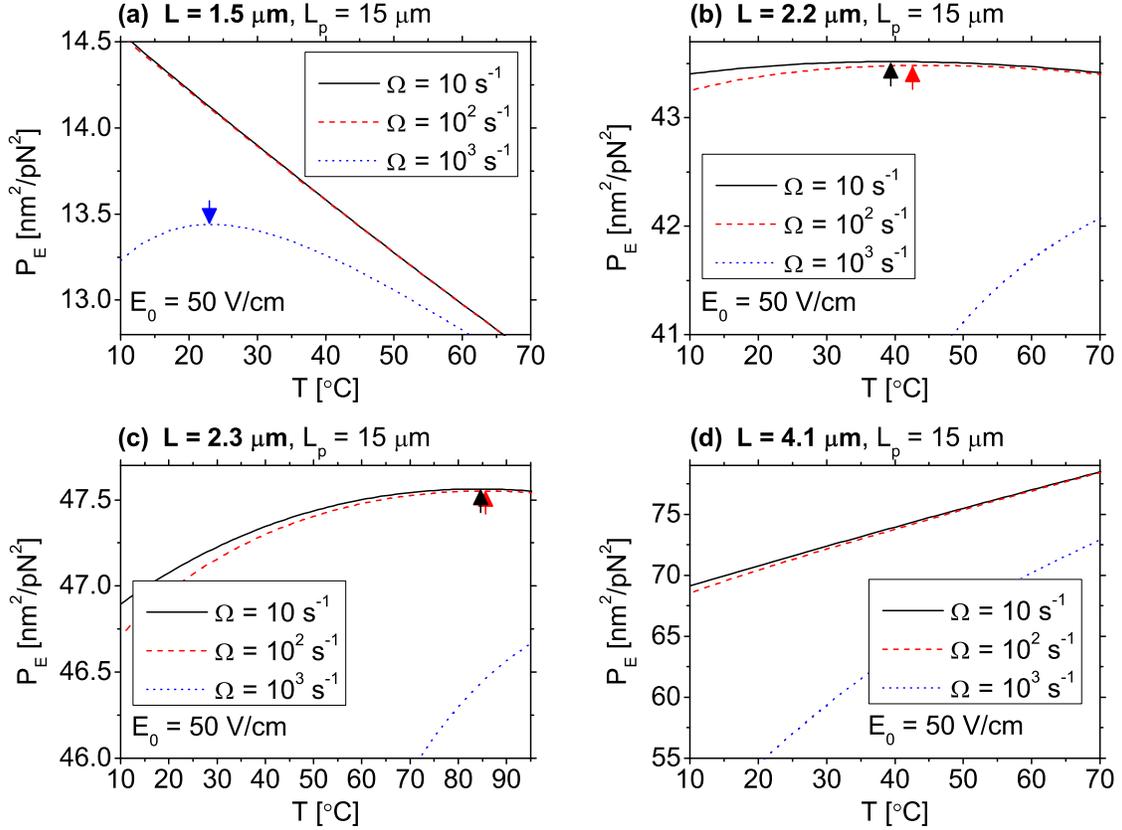}
\caption{\label{fig:5} The power amplification factor $P_{E}$ vs. temperature in Celsius for three frequencies $\Omega=10$ s$^{-1}$ (solid), $\Omega=10^{2}$ s$^{-1}$ (dashed) and $\Omega=10^{3}$ s$^{-1}$ (dotted), with HI included. For the stiff chain taking parameters of a F-actin filament with $a=6$ nm, $L_{p}=15 ~\mu$m and $\lambda=-1$ e/nm, the $P_{E}$ can have maxima at optimal temperatures indicated by the arrows for contour lengths, (a)$L=1.5~\mu$m, (b)$L=2.2~\mu$m, (c)$L=2.3~\mu$m and (d)$L=4.1~\mu$m.}
\end{center}
\end{figure*}
Under an electric field $E_{0}=50$ V/cm, a tethered stiff chain with parameters of F-actin filament fragment, $L >1~\mu$m, $L_{p}=15~\mu$m, $\lambda=-1$ e/nm, and $a=6$ nm, is extended as much as $Z_{0}/L > 0.9$ following the Eq.~(\ref{eq:eteE}).
Considering a small additive driving field $\delta E \sin\left(\Omega t\right)$, we plot the $P_{E}(T)$ for the cases of the stiff chain extension at four different contour lengths, in Fig.~\ref{fig:5}, including the hydrodynamic interactions.
This also shows that the ESR occurs for optimal temperature, similar to the case driven by the stretching force. In the slow driving regime (solid curves in the Fig.~\ref{fig:5}(a), (b), (c) and (d)), the $P_{E}$ also show distinct behaviors within the temperature considered, which are monotonically decreasing, non-monotonically varying with a peak, and monotonically increasing, respectively.

Using the single relaxation time approximation $\tilde{P}_{E}=\left({\langle (\delta Z)^{2}\rangle_{0}}/{2 k_{B}T}\right)^2 /({1+(\tau\Omega)^2})$.
At low frequencies, temperature peaks are due to by the static susceptibilities.
Following the argument behind Eq.~(\ref{eq:27}), the optimal temperature is obtained via the replacement $f_{0}$ by $L|\lambda|E_{0}/4$ in Eq.~(\ref{eq:27}), as $T^{\ast}=(9/256){|\lambda|E_{0}}{L^3}/(4{k_{B}}{L_p})$, which is consistent with the dependence of $T^{\ast}$ on $L$, in Fig. {\ref{fig:5}} for low frequencies. 

\begin{figure}
\begin{center}
\includegraphics[width=8.6cm]{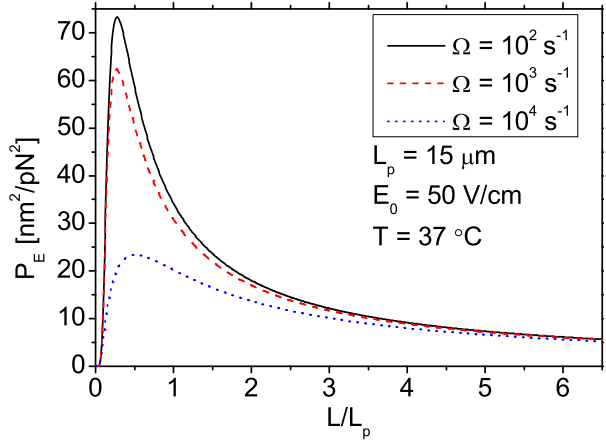}
\caption{\label{fig:6} The $P_{E}$ vs. $L/L_{p}$ with the various $\Omega$ at $T=37^{\circ}$C, with HI included. The $P_{E}$ has the peak with respect to a contour length, which becomes sharper for smaller $\Omega$. All the other parameters are the same as those used in the Fig.~{\ref{fig:5}}.}
\end{center}
\end{figure}
Remarkably, as depicted in Fig.~\ref{fig:6} for a fixed $T$, there exists a sharp peak of the power amplification $P_{E}(L/L_{p})$ at an optimal contour length, which is a new feature of the ESR.
This is due to the competition between the undulational flexibility and the electric-field induced stretching rigidity. As the chain flexibility increases with contour length, the $P_{E}$ increases as in the end-tethered case, while it decreases due to enhanced chain rigidity induced by the electric field effect on the chain segments.
As the angular driving frequency $\Omega$ decreases, the ESR (magnitude of the peak) sharply increases. Note that, in the Fig.~\ref{fig:6}, the optimal contour length (for the peak) is around $4.1~\mu$m, seemingly independent of the angular frequencies.
The replacement of $f_{0}$ by $L|\lambda|E_{0}/4$ in Eq.~(\ref{eq:26}) yields the optimal contour length for the SR peak
\begin{equation}
L^{\ast}=4\left(\frac{16k_{B}T L_{p}}{|\lambda|E_{0}}\right)^{1/3}.
\label{eq:28}
\end{equation}
For the parameters considered therein, $L^{\ast}\approx4~\mu$m in agreement with the Fig.~\ref{fig:6} for the $\Omega \ll \tau^{-1}$.
For this optimal length, the Fig.~\ref{fig:5}(d) shows $P_E(T)$ increases monotonically with $T$.

\begin{figure*}
\begin{center}
\includegraphics[width=16.5cm]{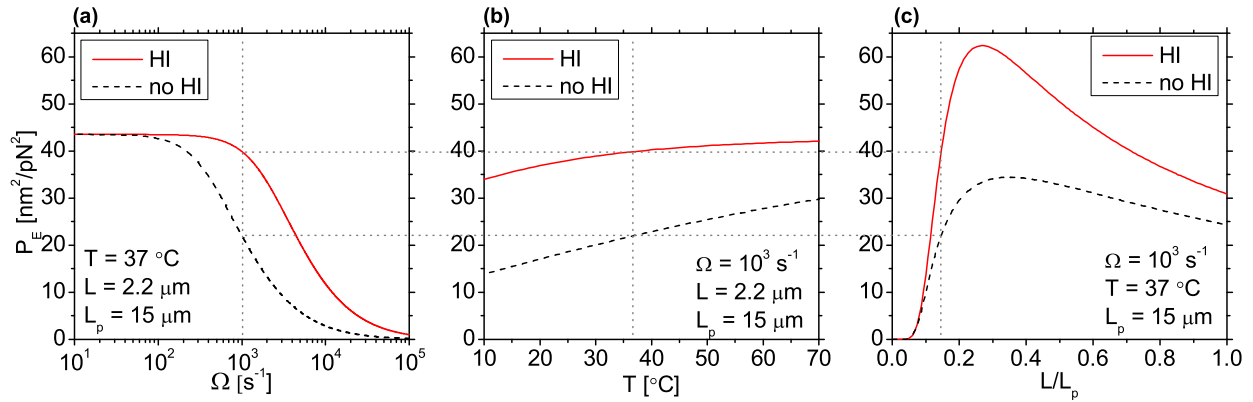}
\caption{\label{fig:7} Enhancement of the power amplification $P_E$ by HI. (a) $P_E$ vs. $\Omega$ with (solid curve) and without HI (dashed curve), at $T=37^{\circ}$C and $L=2.2~\mu$m. HI can enhance the amplification by almost twice at $\Omega=10^{3}$ s$^{-1}$ (see dotted vertical lines) (b) $P_E$ vs. $T$ with and without HI, at $\Omega=10^{3}$ s$^{-1}$ and $L=2.2~\mu$m. (c) $P_E$ vs. $L/L_{p}$ with and without HI, at $T=37^{\circ}$C and $\Omega=10^{3}$ s$^{-1}$. Each cross point of the dotted lines depicts the $P_E$ at $\Omega=10^{3}$ s$^{-1}$, $T=37^{\circ}$C and $L=2.2~\mu$m. All the other parameters are the same as those used in the Fig.~{\ref{fig:5}}.}
\end{center}
\end{figure*}
In Fig.~\ref{fig:7}, we also show HI effect on the $P_E$ for the field-driven case, where its dependency on $\Omega$ (or $T$) is similar to the tension-driven case.
We find that the magnitude of $P_{E}(\Omega)$ at $T=37^{\circ}$C and $L=2.2~\mu$m can be enhanced by almost twice that at $\Omega > \tau^{-1} \simeq 10^{2}$ s$^{-1}$, due to HI, as depicted in Fig.~\ref{fig:7}(a).
Fig.~\ref{fig:7}(b) and Fig.~\ref{fig:7}(c) depict $P_E$ vs. $T$ and $P$ vs. $L/L_{p}$ with and without HI, respectively for the parameters indicated.
Remarkably, as shown in the Fig.~\ref{fig:7}(c), the ESR peak can be highly magnified by HI at the optimal length.

\section{The effective 1-D dynamics of semi-flexible chain extension and entropic stochastic resonance}

Conventionally, theoretical study of SR has mostly considered one (or few) dimensional stochastic system, for example, a Brownian particle under a one-dimensional (1D) double well potential.
On the other hand, hitherto, we have investigated the SR of a stretched chain by considering the motion of many segments, which are locally governed by the elastic interactions under condition of the segmental inextensibility as well as the hydrodynamic interaction.
To understand entropic and dissipative nature of SR, we construct the effective 1-D dynamics of the chain extension $Z(t)$, based on the single relaxation time approximation.

\begin{figure}
\begin{center}
\includegraphics[width=8.6cm]{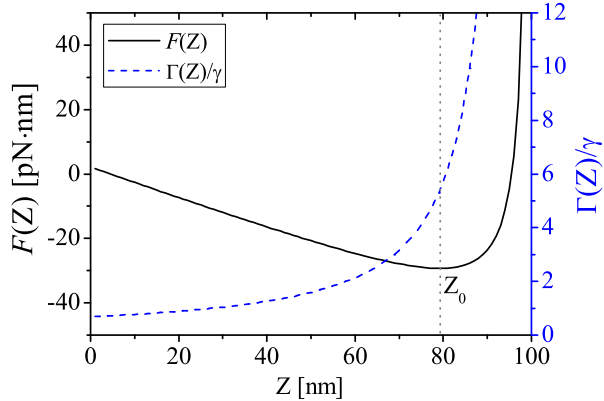}
\caption{\label{fig:free} The free energy $\mathcal{F}(Z)$ (solid curve) and effective friction coefficient $\Gamma(Z)$ divided by $\gamma=6\pi\eta(a/2)$ (dashed curve) as functions of the extension $Z$. The dotted vertical line depicts $Z_{0}$ ($\approx 80$ nm) where free energy is the minimum. We use $T=37^{\circ}$C, $L=100$ nm, $L_{p}=50$ nm, $a=2$ nm and $f_{0}=0.5$ pN.}
\end{center}
\end{figure}

First the Eq.~(\ref{eq:10}) allows us to derive an effective potential or free energy of the chain extended by the constant tension $f_{0}$,
\begin{eqnarray}
\mathcal{F}(Z)&=&\frac{k_{B}T}{4L_{p}}\frac{L}{1-Z/L}-f_{0}Z\nonumber\\
&=&\frac{k_{B}T}{4L_{p}}\left(\frac{L}{1-Z/L}-\frac{Z}{(1-Z_{0}/L)^{2}}\right).
\label{eq20}
\end{eqnarray}
Here $Z_{0}=\langle Z(0)\rangle$ is the equilibrium extension where the free energy is minimum. 
The $\mathcal{F}(Z)$ as depicted in Fig.~\ref{fig:free}, is anharmonic, rising rapidly as the chain is fully extended, due to the chain entropy (the first term in Eq.~(\ref{eq20})).

The 1-D chain dynamics we consider is essentially that of the Brownian motion fluctuating around the free energy minimum. Using the approximation of a single relaxation time, we now establish the effective dynamical equation of the slow chain extension.
It is assumed to be the Langevin equation of the form,
\begin{equation}
\Gamma\dot{Z}=-\frac{\partial \mathcal{F}(Z)}{\partial Z}+\delta f(t)+\xi(Z,t).
\label{eq:L2}
\end{equation}
Here $\Gamma$ is the hydrodynamic drag coefficient of the whole chain under extension $Z$. 
It is related to the random force $\xi(Z,t)$ via a fluctuation-dissipation theorem,
\begin{equation}
\langle\xi(Z,t)\xi(Z,0)\rangle=2k_{B}T\Gamma\delta(t).
\end{equation}
Since, in our approximation, $\langle Z(t)\rangle$ relaxes exponentially toward $Z_{0}$ with the characteristic time $\tau$ in the absence of $\delta f(t)$, we have
\begin{equation}
\dot{Z}=-\frac{1}{\tau}(Z-Z_{0})+\frac{\xi(Z,t)}{\Gamma(Z)}.
\label{eq:stoeq}
\end{equation}
Matching the above with Eq.~(\ref{eq:L2}) allows us to find the $Z$-dependent drag,
\begin{eqnarray}
\Gamma(Z)&=&\frac{\tau}{Z-L(1-\frac{k_{B}T}{\sqrt{4\kappa f_{0}}})}\left[\frac{k_{B}T}{4L_{p}}\frac{1}{(1-Z/L)^{2}}-f_{0}\right]\nonumber\\
&=&\frac{\tau}{Z-Z_{0}}\frac{k_{B}T}{4L_{p}}\left[\frac{1}{(1-Z/L)^{2}}-\frac{1}{(1-Z_{0}/L)^{2}}\right].\nonumber\\
\label{eq:22}
\end{eqnarray}
In the Fig.~\ref{fig:free}, $\Gamma(Z)/\gamma$ is depicted by the dashed curve with segmental friction $\gamma$ put as $6\pi\eta(a/2)$.
$\Gamma(Z)$ rapidly increases with $Z$ above $Z_0 \approx 80$ nm as depicted by the dotted line. This is mostly due to effect of the chain entropy that prohibits the chain to extend to the full length, in this effective model.
It should be noted also that HI manifests through $\tau$, Eq.~(\ref{eq:longtau}), by making it, consequently the chain drag (Eq.~(\ref{eq:22})) smaller by the factor $\sim K_{0}(\pi a/L)\approx |\ln(\pi a/L)|$ compared with that without HI.
Consequently, the chain extension dynamics, effectively reduced to a single degree of freedom $Z$, is stochastic subject to a anharmonic potential (non-linear force) and a non-constant friction due to the chain entropy and hydrodynamic interactions of the undulating, extended chain.
The extension oscillates around $Z_{0}$ in maximum coherence to an oscillating force at SR condition.
Compared with the SR conventionally studied in 1D bistable systems with a constant friction, there are two points worthy of remarks: First, the monostable free energy arising the chain connectivity and entropy manifest the SR (ESR). Second, another feature of the cooperativity, the hydrodynamic interaction, can enhance it by reducing the drag.

\section{summary and discussion}

We analytically have studied dynamical response and associated stochastic resonance (SR) in a stretched semi-flexible chain to an oscillatory (AC) field or force.
Incorporating the chain segmental elasticity and inextensibility, we set up an effective Hamiltonian of the chain in terms of the chain transverse undulation field which is assumed to be small, and evaluated the undulation and extension in response to the external oscillation.
From the dynamic susceptibility of the chain extension, we directly obtained power amplification factor as a measure of the SR.
The chain extension is then found to be most synchronous and coherent to the oscillation at an optimal noise-strength (temperature), revealing an SR of entropic nature.
In particular, for the case of an applied AC field we found a new type of the entropic SR (ESR), where the power amplification is maximum at an optimal chain length due to competition between associated conformational entropy (chain flexibility) and electric field (chain rigidity).
This ESR is enhanced by the hydrodynamic interactions between the chain segments, due to its cooperativity even in overdamping media.

On the other hand, from the way the chain extension responds to a time-dependent field or force, we have constructed the time correlation function of the stretched chain extension in its absence, using the linear response theory.
It decays more rapidly than the exponential decay with the slowest relaxation of time associated with the transverse undulation of longest wave-length.
Yet the latter describes quite accurately the long time behavior of the stretched chain dynamics and the SR to low frequency oscillation.
We have constructed the relevant 1-D dynamics of the stretched chain extension, which is found to be subject to a monostable free energy due to entropic force and a drag reduced by the hydrodynamic interaction, both of which depend upon the chain extension.
This means that stretched polymer is a non-linear monostable system to support a novel kind of ESR.
The simple 1-D stochastic equation we derived (Eq.~(\ref{eq:stoeq})) should be useful to simulate the stretched semi-flexible chain dynamics in correlation with experiments.

The biopolymers in vivo are usually subject not only to thermal noise (temperature) but also to athermal, non-equilibrium noises of various kind due to active cellular environments.
A non-equilibrium noise is usually characterized by a broadband frequencies, but as we have shown, the component corresponding to the lowest frequency is most amplified at an optimal temperature or an optimal contour length.
Although our semi-flexible chain model may be too simple to represent the real biopolymers in action, our result on SR may suggest biological systems and their environments have evolved to adopt the physiological temperatures for optimal biological functions, as recently have been shown in the SR of ion-channel \cite{parc}.
The novel features of ESR for a single biopolymer we discussed here can be tested by single molecule experiments for stretched biopolymers such as tethered dsDNA \cite{doi:10.1021/la200433r} and stiff actin-flaments \cite{biophysj.107.114538} under oscillating fields.

\begin{acknowledgments}
This work was supported by Brain-Korea 21 program and Korea Research Foundation administered by Ministry of Education and Science,
S. Korea. We acknowledge valuable discussions with Fabio Marchesoni.
\end{acknowledgments}

\newpage 

\begin{thebibliography}{21}
\expandafter\ifx\csname natexlab\endcsname\relax\def\natexlab#1{#1}\fi
\expandafter\ifx\csname bibnamefont\endcsname\relax
  \def\bibnamefont#1{#1}\fi
\expandafter\ifx\csname bibfnamefont\endcsname\relax
  \def\bibfnamefont#1{#1}\fi
\expandafter\ifx\csname citenamefont\endcsname\relax
  \def\citenamefont#1{#1}\fi
\expandafter\ifx\csname url\endcsname\relax
  \def\url#1{\texttt{#1}}\fi
\expandafter\ifx\csname urlprefix\endcsname\relax\def\urlprefix{URL }\fi
\providecommand{\bibinfo}[2]{#2}
\providecommand{\eprint}[2][]{\url{#2}}

\bibitem[{\citenamefont{Luger et~al.}(1997)\citenamefont{Luger, Mader,
  Richmond, Sargent, and Richmond}}]{luger}
\bibinfo{author}{\bibfnamefont{K.}~\bibnamefont{Luger}},
  \bibinfo{author}{\bibfnamefont{A.~W.} \bibnamefont{Mader}},
  \bibinfo{author}{\bibfnamefont{R.~K.} \bibnamefont{Richmond}},
  \bibinfo{author}{\bibfnamefont{D.~F.} \bibnamefont{Sargent}},
  \bibnamefont{and} \bibinfo{author}{\bibfnamefont{T.~J.}
  \bibnamefont{Richmond}}, \bibinfo{journal}{Nature}
  \textbf{\bibinfo{volume}{389}}, \bibinfo{pages}{251} (\bibinfo{year}{1997}).

\bibitem[{\citenamefont{Bulsara and Gammaitoni}(1996)}]{sr1}
\bibinfo{author}{\bibfnamefont{A.}~\bibnamefont{Bulsara}} \bibnamefont{and}
  \bibinfo{author}{\bibfnamefont{L.}~\bibnamefont{Gammaitoni}},
  \bibinfo{journal}{Physics Today} \textbf{\bibinfo{volume}{49}},
  \bibinfo{pages}{39} (\bibinfo{year}{1996}).

\bibitem[{\citenamefont{Gammaitoni et~al.}(1998)\citenamefont{Gammaitoni,
  H\"{a}nggi, Jung, and Marchesoni}}]{sr2}
\bibinfo{author}{\bibfnamefont{L.}~\bibnamefont{Gammaitoni}},
  \bibinfo{author}{\bibfnamefont{P.}~\bibnamefont{H\"{a}nggi}},
  \bibinfo{author}{\bibfnamefont{P.}~\bibnamefont{Jung}}, \bibnamefont{and}
  \bibinfo{author}{\bibfnamefont{F.}~\bibnamefont{Marchesoni}},
  \bibinfo{journal}{Rev. Mod. Phys.} \textbf{\bibinfo{volume}{70}},
  \bibinfo{pages}{223} (\bibinfo{year}{1998}).

\bibitem[{\citenamefont{Evstigneev et~al.}(2004)\citenamefont{Evstigneev,
  Reimann, Pankov, and Prince}}]{0295-5075-65-1-007}
\bibinfo{author}{\bibfnamefont{M.}~\bibnamefont{Evstigneev}},
  \bibinfo{author}{\bibfnamefont{P.}~\bibnamefont{Reimann}},
  \bibinfo{author}{\bibfnamefont{V.}~\bibnamefont{Pankov}}, \bibnamefont{and}
  \bibinfo{author}{\bibfnamefont{R.~H.} \bibnamefont{Prince}},
  \bibinfo{journal}{Europhys. Lett.} \textbf{\bibinfo{volume}{65}},
  \bibinfo{pages}{7} (\bibinfo{year}{2004}).

\bibitem[{\citenamefont{Xiang-Dong et~al.}(2009)\citenamefont{Xiang-Dong, Feng,
  and Yu-Rong}}]{0253-6102-51-2-19}
\bibinfo{author}{\bibfnamefont{L.}~\bibnamefont{Xiang-Dong}},
  \bibinfo{author}{\bibfnamefont{G.}~\bibnamefont{Feng}}, \bibnamefont{and}
  \bibinfo{author}{\bibfnamefont{Z.}~\bibnamefont{Yu-Rong}},
  \bibinfo{journal}{Comm. Theoret. Phys.}
  \textbf{\bibinfo{volume}{51}}, \bibinfo{pages}{283} (\bibinfo{year}{2009}).

\bibitem[{\citenamefont{Asfaw and Sung}(2010)}]{asfaw}
\bibinfo{author}{\bibfnamefont{M.}~\bibnamefont{Asfaw}} \bibnamefont{and}
  \bibinfo{author}{\bibfnamefont{W.}~\bibnamefont{Sung}},
  \bibinfo{journal}{Europhys. Lett.} \textbf{\bibinfo{volume}{90}},
  \bibinfo{pages}{30008} (\bibinfo{year}{2010}).

\bibitem[{\citenamefont{Lee and Sung}(2001)}]{PhysRevE.63.021115}
\bibinfo{author}{\bibfnamefont{S.}~\bibnamefont{Lee}} \bibnamefont{and}
  \bibinfo{author}{\bibfnamefont{W.}~\bibnamefont{Sung}},
  \bibinfo{journal}{Phys. Rev. E} \textbf{\bibinfo{volume}{63}},
  \bibinfo{pages}{021115} (\bibinfo{year}{2001}),

\bibitem[{\citenamefont{Granek}(2004)}]{granekprl}
\bibinfo{author}{\bibfnamefont{Y.}~\bibnamefont{B.}~\bibnamefont{Raviv}},
\bibinfo{author}{\bibfnamefont{W.}~\bibnamefont{Z.}~\bibnamefont{Zhao}},
\bibinfo{author}{\bibfnamefont{C.}~\bibnamefont{H.}~\bibnamefont{Wiggins}} \bibnamefont{and}
\bibinfo{author}{\bibfnamefont{R.}~\bibnamefont{Granek}},
  \bibinfo{journal}{Phys. Rev. Lett.
} \textbf{\bibinfo{volume}{92}}, \bibinfo{pages}{098101}
  (\bibinfo{year}{2004}).

\bibitem[{\citenamefont{Hiraiwa and Ohta}(2009)}]{ohta}
\bibinfo{author}{\bibfnamefont{T.}~\bibnamefont{Hiraiwa}} \bibnamefont{and}
  \bibinfo{author}{\bibfnamefont{T.}~\bibnamefont{Ohta}},
  \bibinfo{journal}{Macromolecules} \textbf{\bibinfo{volume}{42}},
  \bibinfo{pages}{7553} (\bibinfo{year}{2009}).

\bibitem[{\citenamefont{Schulz}(1993)}]{PhysRevLett.71.1864}
\bibinfo{author}{\bibfnamefont{H.~J.} \bibnamefont{Schulz}},
  \bibinfo{journal}{Phys. Rev. Lett.} \textbf{\bibinfo{volume}{71}},
  \bibinfo{pages}{1864} (\bibinfo{year}{1993}).

\bibitem[{\citenamefont{Marko and Siggia}(1995)}]{marko}
\bibinfo{author}{\bibfnamefont{J.~F.} \bibnamefont{Marko}} \bibnamefont{and}
  \bibinfo{author}{\bibfnamefont{E.~D.} \bibnamefont{Siggia}},
  \bibinfo{journal}{Macromolecules} \textbf{\bibinfo{volume}{28}},
  \bibinfo{pages}{8759} (\bibinfo{year}{1995}).

\bibitem[{\citenamefont{Brochard-Wyart
  et~al.}(1999)\citenamefont{Brochard-Wyart, Buguin, and de~Gennes}}]{epl47}
\bibinfo{author}{\bibfnamefont{F.}~\bibnamefont{Brochard-Wyart}},
  \bibinfo{author}{\bibfnamefont{A.}~\bibnamefont{Buguin}}, \bibnamefont{and}
  \bibinfo{author}{\bibfnamefont{P.~G.} \bibnamefont{de~Gennes}},
  \bibinfo{journal}{Europhys. Lett.} \textbf{\bibinfo{volume}{47}},
  \bibinfo{pages}{171} (\bibinfo{year}{1999}).

\bibitem[{\citenamefont{Korson et~al.}(1969)\citenamefont{Korson, Drost-Hansen,
  and Millero}}]{doi:10.1021/j100721a006}
\bibinfo{author}{\bibfnamefont{L.}~\bibnamefont{Korson}},
  \bibinfo{author}{\bibfnamefont{W.}~\bibnamefont{Drost-Hansen}},
  \bibnamefont{and} \bibinfo{author}{\bibfnamefont{F.~J.}
  \bibnamefont{Millero}}, \bibinfo{journal}{J. Phys. Chem.}
  \textbf{\bibinfo{volume}{73}}, \bibinfo{pages}{34} (\bibinfo{year}{1969}).

\bibitem[{\citenamefont{Benetatos and Frey}(2004)}]{benetatos}
\bibinfo{author}{\bibfnamefont{P.}~\bibnamefont{Benetatos}} \bibnamefont{and}
  \bibinfo{author}{\bibfnamefont{E.}~\bibnamefont{Frey}},
  \bibinfo{journal}{Phys. Rev. E.} \textbf{\bibinfo{volume}{70}},
  \bibinfo{pages}{051806} (\bibinfo{year}{2004}).

\bibitem[{\citenamefont{Maier et~al.}(2002)\citenamefont{Maier, Seifert, and
  Radler}}]{maier}
\bibinfo{author}{\bibfnamefont{B.}~\bibnamefont{Maier}},
  \bibinfo{author}{\bibfnamefont{U.}~\bibnamefont{Seifert}}, \bibnamefont{and}
  \bibinfo{author}{\bibfnamefont{J.~O.} \bibnamefont{Radler}},
  \bibinfo{journal}{Europhys. Lett.} \textbf{\bibinfo{volume}{60}},
  \bibinfo{pages}{622} (\bibinfo{year}{2002}).

\bibitem[{\citenamefont{Hori et~al.}(2007)\citenamefont{Hori, Prasad, and
  Kondev}}]{hori}
\bibinfo{author}{\bibfnamefont{Y.}~\bibnamefont{Hori}},
  \bibinfo{author}{\bibfnamefont{A.}~\bibnamefont{Prasad}}, \bibnamefont{and}
  \bibinfo{author}{\bibfnamefont{J.}~\bibnamefont{Kondev}},
  \bibinfo{journal}{Phys. Rev. E.} \textbf{\bibinfo{volume}{75}},
  \bibinfo{pages}{041904} (\bibinfo{year}{2007}).

\bibitem[{\citenamefont{Odijk}(1977)}]{OSF1}
\bibinfo{author}{\bibfnamefont{T.}~\bibnamefont{Odijk}}, \bibinfo{journal}{J.
  Polym. Sci., Polym. Phys. Ed.} \textbf{\bibinfo{volume}{15}},
  \bibinfo{pages}{477} (\bibinfo{year}{1977}).

\bibitem[{\citenamefont{Skolnick and Fixman}(1977)}]{OSF2}
\bibinfo{author}{\bibfnamefont{J.}~\bibnamefont{Skolnick}} \bibnamefont{and}
  \bibinfo{author}{\bibfnamefont{M.}~\bibnamefont{Fixman}},
  \bibinfo{journal}{Macromolecules} \textbf{\bibinfo{volume}{10}},
  \bibinfo{pages}{944} (\bibinfo{year}{1977}).

\bibitem[{\citenamefont{Parc and Sung}(2009)}]{parc}
\bibinfo{author}{\bibfnamefont{Y.~W.} \bibnamefont{Parc}},
\bibinfo{author}{\bibfnamefont{D.~S.} \bibnamefont{Koh}}, \bibnamefont{and}
  \bibinfo{author}{\bibfnamefont{W.}~\bibnamefont{Sung}},
  \bibinfo{journal}{Eur. Phys. J. B.} \textbf{\bibinfo{volume}{69}},
  \bibinfo{pages}{127} (\bibinfo{year}{2009}).

\bibitem[{\citenamefont{Guan et~al.}(2011)\citenamefont{Guan, Wang, and
  Granick}}]{doi:10.1021/la200433r}
\bibinfo{author}{\bibfnamefont{J.}~\bibnamefont{Guan}},
  \bibinfo{author}{\bibfnamefont{B.}~\bibnamefont{Wang}}, \bibnamefont{and}
  \bibinfo{author}{\bibfnamefont{S.}~\bibnamefont{Granick}},
  \bibinfo{journal}{Langmuir} \textbf{\bibinfo{volume}{27}},
  \bibinfo{pages}{6149} (\bibinfo{year}{2011}).
  
\bibitem[{\citenamefont{Arsenault et~al.}(2007)\citenamefont{Arsenault, Zhao,
  Purohit, Goldman, and Bau}}]{biophysj.107.114538}
\bibinfo{author}{\bibfnamefont{M.~E.} \bibnamefont{Arsenault}},
  \bibinfo{author}{\bibfnamefont{H.}~\bibnamefont{Zhao}},
  \bibinfo{author}{\bibfnamefont{P.~K.} \bibnamefont{Purohit}},
  \bibinfo{author}{\bibfnamefont{Y.~E.} \bibnamefont{Goldman}},
  \bibnamefont{and} \bibinfo{author}{\bibfnamefont{H.~H.} \bibnamefont{Bau}},
  \bibinfo{journal}{Biophys. J.} \textbf{\bibinfo{volume}{93}},
  \bibinfo{pages}{L42} (\bibinfo{year}{2007}).
  
\end{thebibliography}

\end{document}